\documentclass[twocolumn, showpacs]{revtex4}
\usepackage{graphicx}
\usepackage{dcolumn}
\usepackage{bm}
\def\ph2{{\it p}-H$_2$}
\def\od2{{\it o}-D$_2$}
\def\Am3{\AA$^{-3}$}

\begin{document}
\title{Structure, superfluidity, and quantum melting of hydrogen clusters}
\author{Fabio Mezzacapo$^{1}$ and Massimo Boninsegni$^{1,2}$}
\affiliation{$^1$Department of Physics, University of Alberta, Edmonton, Alberta, Canada, T6G 2J1\\
$^2$INFM-BEC, Dipartimento di Fisica, Universita' degli Studi di Trento, Via Sommarive 4, I-38050 Povo, Italy}
\date\today

\begin{abstract}
We present results of a  theoretical study of \ph2 and \od2 clusters at low temperature (0.5 K $\le$ $T$ $\le$ 3.5 K), based on Path Integral Monte Carlo simulations. Clusters of $N\le$ 21 \ph2 molecules are nearly entirely superfluid at $T\le$ 1 K. For 22  $\le$ $N$ $\le$ 30, the superfluid response  displays strong variations with $N$, reflecting structural changes that occur on adding or removing even a single molecule. Some  clusters in this size range display {\it quantum melting}, going from solid- to liquid-like as $T\to 0$. Melting is caused by quantum exchanges of molecules. The largest \ph2 cluster for which a significant superfluid response is observed comprises $N$=27 molecules.  Evidence of a finite superfluid response is presented for  \od2 clusters of size up to $N$=14. Magic numbers are observed, at which both types of clusters feature pronounced  stability. 
\end{abstract}

\maketitle

\section{INTRODUCTION}
\label{intro}
The search for superfluidity (SF) in condensed phases of atomic or molecular species other than Helium is the subject of a significant investigative effort, both from the theoretical and the experimental standpoints. 
Fluids made of light  bosonic molecules such as {\it para}-hydrogen (\ph2)  or {\it ortho}-deuterium (\od2), whose mass is smaller than (or equal to) that of helium atoms, have long been regarded as potential superfluids \cite{ginzburg72}. However, the attractive interaction between two hydrogen molecules results in an equilibrium crystal phase,  with a freezing temperature higher than that at which a superfluid transition might take place \cite{supersolid}. 

Confinement, and reduction of dimensionality, are deemed plausible avenues to stabilize  a metastable liquid phase   down to temperatures sufficiently low that a superfluid transition may be observed. However,  theoretical studies of \ph2 films adsorbed on various substrates \cite{wagner96, nho02, mb04a}, as well as in two mathematical dimensions \cite{mb04b}, have so far yielded  no hint of possible SF. The suggestion was made that SF may occur in a strictly two-dimensional (2D) \ph2 fluid embedded in a crystalline matrix of Alkali atoms \cite{gordillo97}. That prediction was actually based on numerical simulations of model systems of very small size, and there is evidence that the observed superfluid signal is merely a finite-size effect \cite{mb05}. 

On the other hand, sufficiently  small clusters can remain ``liquid-like" at significantly lower $T$ than the bulk, and therefore SF could occur at temperatures that may render its observation possible \cite{noteliquid}. For example, a recently introduced experimental technique, known as Helium NanoDroplet Isolation  spectroscopy (HENDI), allows one to investigate a single molecular impurity embedded in clusters comprising from a few, to several thousand He atoms \cite{scoles92}.  In these experiments, SF of the medium surrounding the molecule (i.e., the cluster) can be inferred by studying the rotational spectrum of the dopant. Specifically, the observed decoupling of the rotation of the molecule from that of the cluster signals the  onset of SF in the cluster.    
Evidence of SF   has been found at $T < 1$ K in helium clusters surrounding a linear carbonyl  sulfide (OCS) impurity, and, with the same technique, in  clusters of $N$=14-16 \ph2 molecules surrounding the same impurity; these clusters are enclosed in a larger  helium droplet \cite{grebenev98, grebenev00}. The same experiment failed to detect superfluid behavior of  \od2 clusters.

These observations constitute the starting point for an intense theoretical effort, aimed at gaining a deeper understanding of SF in clusters of  \ph2 and, possibly, \od2 molecules. Quantum Monte Carlo (QMC) methods appear to be the most powerful theoretical tool to attack such an interesting problem, and have already been successfully  employed to characterize the structure and  SF  of doped $^4$He clusters both at finite and zero temperature \cite{cep03, sav03, me0405, saverio04}. For instance, Path Integral Monte Carlo (PIMC) simulations have yielded evidence of SF in (\ph2)$_N$ clusters  seeded with an OCS molecule [OCS@(\ph2)$_N]$,  for 10  $\le$ $N$ $\le$ 17, as well as  in smaller complexes such as  OCS@(\ph2)$_5$ and OCS@(\od2)$_5$ \cite{kwon02, paesani05, kwon05}. Reptation Quantum Monte Carlo  simulations have provided information about the structure of CO@(\ph2)$_N$, for $N\sim 12$, and have clarified some  features of the experimentally measured infrared spectra 
 \cite{saverio05, saverio05a}. 

Obviously, of crucial importance to the full characterization of the experiments, as well as to the understanding of the microscopic origin of  SF, remains the study of pristine (i.e., undoped) clusters. In this sense, hydrogen is even richer a playground than helium. For example, \ph2/\od2 clusters provide a unique opportunity to study the effect of mass on the superfluid and structural properties of such a small systems, because the \od2 molecular mass is twice that of \ph2, while the intermolecular interaction is very nearly the same.  Moreover, because the equilibrium phase of hydrogen is crystalline at low $T$, studying clusters of molecular hydrogen offers insight in the evolution of the physical properties as the number of particles is increased.

The first theoretical investigation 
of pure \ph2 clusters, based on PIMC simulations, was published in 1991; it 
was restricted to three cluster sizes, namely $N$=13, 18 and 33. Two clusters were found superfluid and ``liquid-like", while (\ph2)$_{33}$ was observed to be ``solid-like" at $T \le$ 2 K \cite{sindzingre91}. 

In a recent article, we carried out a systematic study of the physical properties of \ph2 clusters, as a function of both size  and temperature ($N \le$ 40 and 0.5 K $\le$ $T$ $\le$ 4 K) \cite{noi06}. Such a study was based on PIMC simulations making use of a recently introduced {\it worm} algorithm (WA) \cite{MBworm}.
Specifically, we computed superfluid fraction $\rho_S$ and density profiles, pointing out the remarkable dependence on $N$ of   $\rho_S$, for 22 $\le$ $N$ $\le$ 30 and how, in this size range, some clusters display   coexistence of  insulating (solid-like) and superfluid (liquid-like)  phases, with the latter, induced by zero-point motion and quantum exchanges, becoming dominant as $T$ is lowered. We refer to this peculiar behavior as ``quantum melting".   Although experimental data for pristine clusters are not yet available, novel techniques based on Raman spectroscopy may soon be able to isolate pure clusters and characterize their superfluid properties \cite{toennies04}. 

The purpose of the present paper is twofold; on the one hand, we  provide additional results for \ph2 clusters, chiefly the energetics but also for the superfluid fraction ($\rho_S(T)$) and the structure of the clusters as a function of their size, also compare results obtained with different intermolecular potentials. We also offer an expanded discussion of the material presented in Ref. \onlinecite{noi06}. Moreover, 
we present novel results for  pure \od2 clusters of size up to $N$=20 molecules, down to $T$=0.5 K. We report superfluid behavior of  \od2 clusters of as many as 14 molecules; in these clusters as well, structure and superfluid response are closely related, and non-trivially dependent on $N$. Specific clusters [i.e., (\od2)$_{13}$ and (\od2)$_{19}$]   feature  greater stability, confirming the existence of ``magic" numbers $N$ of molecules.  

The reminder of this manuscript is organized as follows: in the next section, we describe our microscopic model Hamiltonian and offer basic details of our simulations (the reader is referred to Refs. \onlinecite{MBworm} for a full description of the WA).  In Sec \ref{results} we present and discuss our findings. In particular,  we provide additional details on our results previously published in Ref. \onlinecite{noi06} and illustrate our new numerical data for both \ph2 and \od2 pure clusters. Finally, we outline the main physical conclusions of our work.

\section{METHODOLOGY}
\label{method}
We model the system of our interest as a collection of $N$ \ph2 (\od2)  molecules, regarded as point particles. The quantum mechanical hamiltonian is the following: 
\begin{equation}
H=-\lambda \sum_i^N \nabla_i^2 + V(\mathbf{R})
\end{equation}
where $\lambda$=12.031 K\AA$^2$ for \ph2 (half this value for \od2); $\mathbf{R}\equiv  \mathbf{r}_1,\mathbf{r}_2\cdots\mathbf{r}_N$ is a collective coordinate referring to all $N$ particles in the system, and $V(\mathbf{R})$ the total potential energy of the configuration $\mathbf{R}$.  

In most of the published numerical work on clusters of molecular hydrogen,  the potential energy $V$ is expressed as a sum of pair-wise contributions, each represented by  a spherically symmetric potential; this work is no exception, in this regard.
For the majority of our calculations, we have made use of the Silvera-Goldman (SG) pair potential \cite{SG}, which is the most commonly employed in previous investigations.  For comparison purposes, however, we have also obtained results with a different intermolecular potential,  due to Buck \cite{buck83,buck84,note}.  

Our simulations are carried out by means of a {\it canonical} variant of the WA presented in Refs. \onlinecite{MBworm}. The only difference, with respect to the fully {\it grand canonical} implementation, is that all the so-called $Z$-sector  configurations (which are utilized to compute thermodynamic expectation values) have the same number of particles (i.e., there are no fluctuations of $N$); this proves more convenient than a fully grand canonical implementation, as we are interested in characterizing properties of clusters as a function of $N$. As in any PIMC simulation, a high-temperature approximation for the many-body density matrix is required, and we have used one that is accurate up to fourth order \cite{voth} in the imaginary time step $\epsilon$; we empirically found  the value $\epsilon={1}/{640}$ K$^{-1}$ to yield estimates indistinguishable, within our quoted statistical uncertainties, from those extrapolated to the $\epsilon\to 0$ limit (i.e., the limit where the method becomes formally {\it exact}). We used in our calculations a cubic box of 50 \AA\ side, with periodic boundary conditions (PBC) \cite{notepbc}.  In previous calculations (e.g., Ref. \onlinecite{sindzingre91}), an artificial confining potential was adopted, in order to prevent molecules from evaporating (i.e., to keep the clusters together). We found this device unnecessary, as  all the clusters studied here stay together, with no sign of evaporation, without such an external potential. 

As mentioned above, in the present work we computed  energetics,   radial density profiles and the superfluid fraction $\rho_S(T)$ of \ph2 and \od2 clusters. All of the estimates reported here,  as well as the physical effects discussed in the next section, are observed to be independent of the initial configuration of the clusters. 

 In a finite system, and within the PIMC formalism, it can be shown \cite{Cephe} that  $\rho_S(T)$, defined as the fraction of the system that decouples from an externally induced rotation,  is expressed by:
\begin{equation}
{\rho_S}(T)=\frac{4m^2T}{\hbar^2 I_c}\langle A^2\rangle
\end{equation}
Here, $\langle...\rangle$ stands for thermal average; $I_c$ is the classical moment of inertia of the cluster, and $A$ the total area swept by the many-particle paths,  projected onto a plane perpendicular to one of the three equivalent rotation axes. 

\section{RESULTS}
In this section we present our results for pure \ph2 and \od2 clusters.  Some of the results shown for the superfluid density of \ph2 clusters have appeared in Ref. \cite{noi06}.  The results for \od2 clusters are presented in the last subsection, in which we also point out differences and similarities between \ph2 and \od2 clusters.   
\label{results}
\subsection{\ph2 clusters}
\label{ph2results}

\subsubsection{Energetics}
\begin{figure}[t]
\includegraphics[scale=0.32, angle=-90]{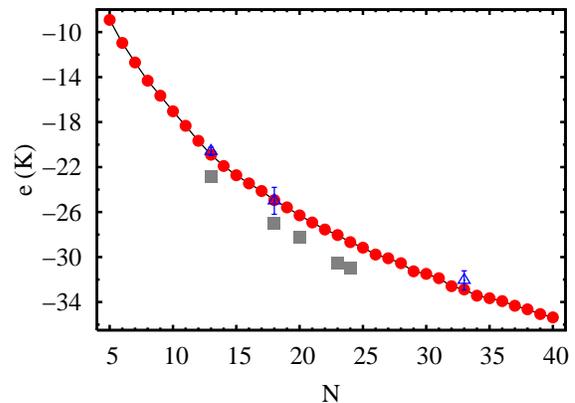}
\caption{(color online). Energy per \ph2 molecule versus cluster size $N$, at $T$ = 1 K. Estimates have been obtained using the SG (circles) and the Buck (boxes) potential. When not shown, statistical errors are of the order of, or smaller than the symbol size. Solid line is only a guide to the eye. Also shown for comparison are results from Ref. \cite{sindzingre91} (triangles).}
\label{energyph2_1K}
\end{figure}

The total energy per \ph2 molecule $e(N)$ as a function of the cluster size $N$, at $T$=1 K, computed using the SG potential, is shown in Fig \ref{energyph2_1K}. Also shown in the figure are the estimates from Ref. \onlinecite{sindzingre91}. For comparison purposes,  we also furnish energetics based on  the Buck potential; in this case, binding energies are considerably greater than those computed with the SG interaction. For example,  a cluster of 23 molecules is characterized by an energy  some 2.5 K more negative than that obtained with the SG potential. Such a difference can be simply ascribed
to the deeper attractive well (roughly 2 K at the minimum) of the Buck potential.

The energy values shown in Fig. \ref{energyph2_1K} for the Buck potential are generally consistent with those reported in a recent paper by Guardiola and Navarro \cite{guardiola06}, who computed them at $T$=0 by means of Diffusion Monte Carlo simulations \cite{notextra}. However, for some clusters our energies are significantly lower than those of Ref. \onlinecite{guardiola06}. For example, for $N$=23 the value quoted in Ref. \onlinecite{guardiola06} is $-29.94(2)$ K, whereas our own value at $T$=1 K is $-30.513(15)$ K.  
At present, the reason for such a discrepancy is unknown. We note, however, that our data for the Buck potential are in agreement with those obtained independently by other authors, using $T$=0 QMC methods \cite{javier06,dorte}.

The energy per \ph2 molecule is  a monotonically decreasing function of the cluster size; the simple formula 
\begin{equation}
e(N)=A_V+A_SN^{-\frac{1}{3}}+A_CN^{-\frac{2}{3}}
\label{fit}
\end{equation}
where $A_V$, $A_S$ and $A_C$ are the well-known volume, surface and curvature terms, fails to yield
a reasonable fit to our data, as well as an accurate extrapolation of the bulk energy ($\sim -90$ K), to indicate that the physics of the bulk is not yet approached by clusters of size studied here. This should be compared with the case of small $^4$He clusters (of size up to $N$=50),
for which the same formula provides an acceptable fit to the data, even though the extrapolated  bulk energy differs from the actual value by over 0.5 K \cite{toennies06}. 

Because in the case of helium, the equilibrium condensed phase is liquid,  no qualitative structural change occurs as the size of the cluster is increased. For hydrogen, on the other hand, the structure must evolve from liquid- to solid-like, presumably in a non-monotonic fashion, but rather going through structures of different shapes and geometries. It is therefore scarcely surprising that a simple formula like (\ref {fit}) should not offer equally acceptable a fit as for helium clusters.

\begin{figure}[t]
\includegraphics[scale=0.32, angle=-90]{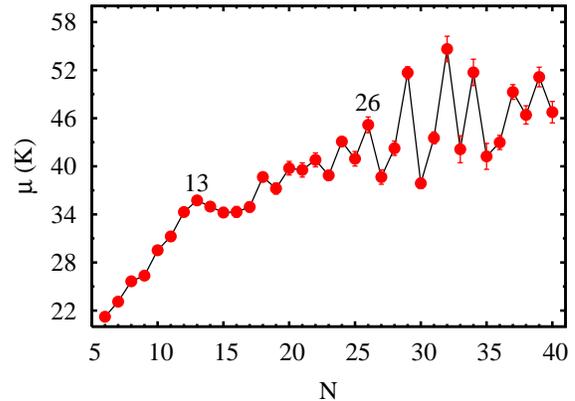}
\caption{(color online). Chemical potential of \ph2 clusters versus cluster size $N$. When not shown, statistical errors are of the order of, or smaller than the symbol size. Solid line is only a guide to the eye. Numbers refer to stable (``magic") clusters.}
\label{cmptlph2}
\end{figure}

Recent $T$=0 QMC calculations  \cite{guardiola06, javier06} have pointed out the existence of magic sizes (i.e., $N$=13) corresponding to particularly stable  clusters. In order to have a clear visual identification of these magic sizes, we plot in Fig. \ref{cmptlph2} the chemical potential $\mu$ as a function of the cluster size defined as  \cite{notextra}:

\begin{equation}
\mu(N)=E(N-1)-E(N)
\label{mu}
\end{equation}
where $E(N)$ is the total energy of a cluster of $N$ molecules.

The chemical potential increases monotonically for $N \le$ 13; for clusters of greater size, its behavior is rather irregular. Local maxima are observed for particular values of $N$ (i.e., $N$=13 and 26). The corresponding clusters are characterized by considerable stability, and rather compact structure. 
 
In the following, we show that \ph2 clusters of size $N <$ 22 display properties consistent with a liquid-like superfluid character;  the stability of (\ph2)$_{13}$ is, in our view, not the sign of the occurrence of a particular solid-like structure, as proposed in Ref. \cite{guardiola06}, but rather of first shell completion (which does not rule out liquid-like behavior) \cite{toennies04,saverio05}. On the other hand, as clarified below, the peak of the chemical potential observed at $N$=26 reflects the distinct solid-like behavior of the cluster (\ph2)$_{26}$.
 
\subsubsection{Superfluidity}
\begin{figure}[]
\includegraphics[scale=0.32, angle=-90]{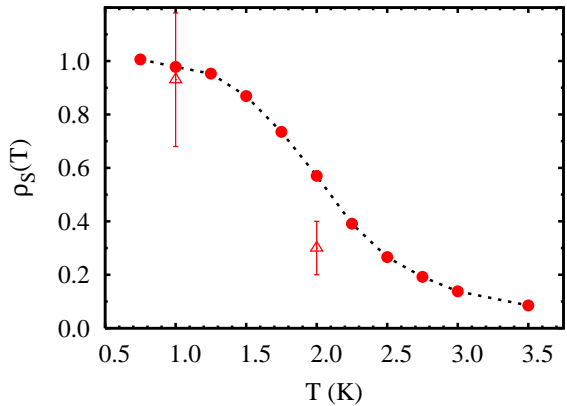}
\caption{(color online). Temperature dependence of the superfluid fraction $\rho_S(T)$ for a cluster of 18 \ph2 molecules (circles). Dotted line is only a guide  to the eye. When not shown, statistical errors are smaller than the symbol size. Also shown for comparison are results from Ref. \cite{sindzingre91} (triangles). }
\label{18_T}
\end{figure}

Data for the superfluid fraction $\rho_S(T)$ of (\ph2)$_{18}$ are shown in Fig \ref{18_T}.  As expected, $\rho_S(T)$ is a monotonic decreasing function of $T$;  this cluster is essentially entirely superfluid (i.e., $\rho_S(T) \simeq$ 1) at $T  \leq$ 1 K. Then, $\rho_S(T)$ drops to a value $\sim$ 0.27    in a temperature interval of  1.5 K and decreases more slowly at higher $T$.  

A finite system cannot undergo a  phase transition in a strict sense; therefore, in order to assign a ``transition temperature'', one must resort to some arbitrary criterion. Because the notion of superfluid fraction is scarcely meaningful when the average number of molecules in the superfluid phase is $\sim$ 1,  we empirically define our ``transition temperature" $T_{\rm c}$ as that at which $N\rho_S(T_{\rm c}) \simeq 2$. Hence the estimated transition temperature is $T_{\rm c} \approx$ 3 K. 

\begin{figure}
\centerline{\includegraphics[scale=0.32, angle=-90]{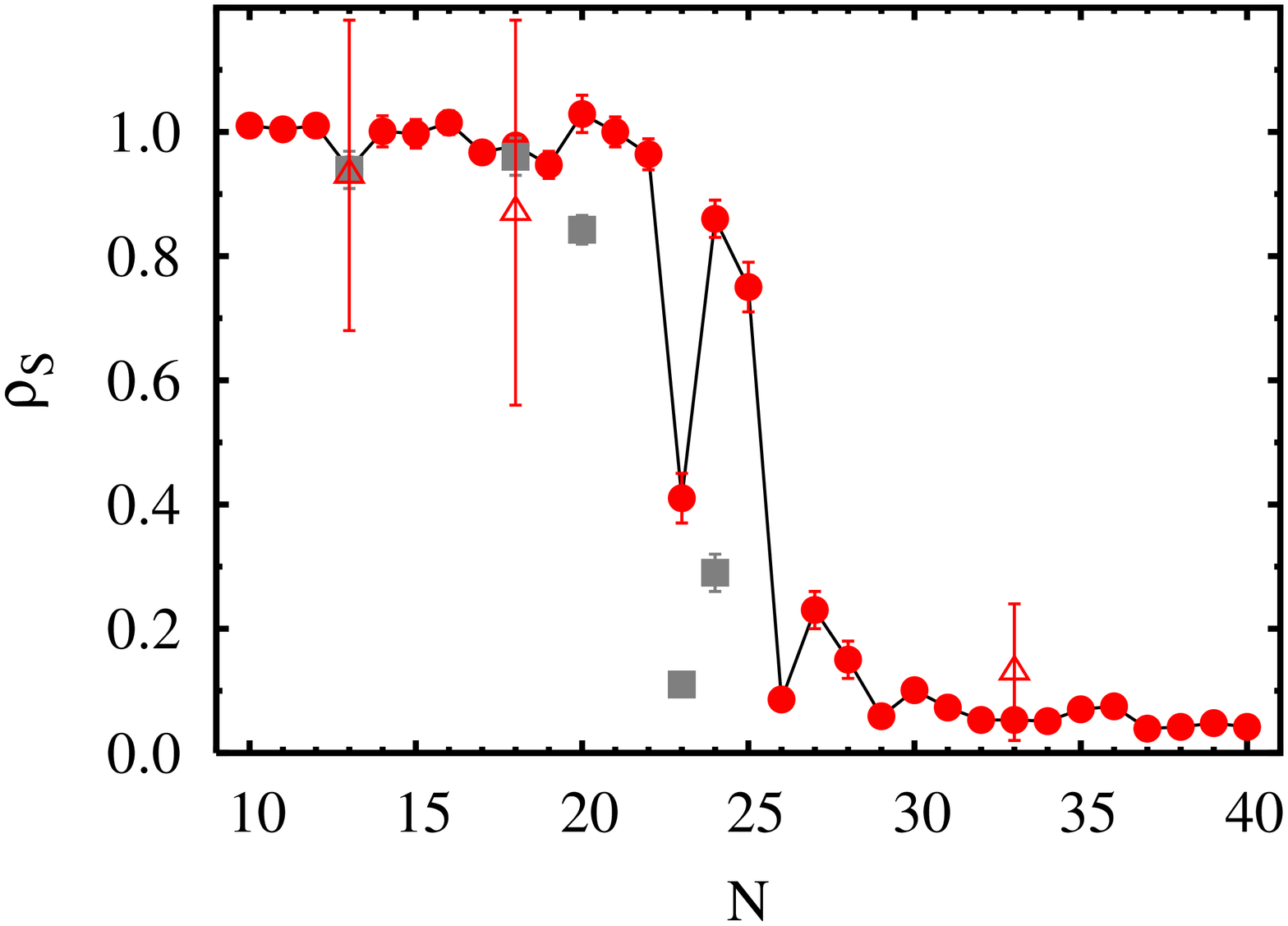}}
\caption{(color online). Superfluid fraction of \ph2 clusters versus cluster size $N$, at $T$=1 K. Estimates have been obtained using the SG (circles) and the Buck (boxes) potential. When not shown, statistical errors are of the order of, or smaller than the symbol size. Solid line is only a guide to the eye. Also shown for comparison are results from Ref. \onlinecite{sindzingre91} (triangles).}
\label{rhosph2_1}
\end{figure}

The superfluid fraction computed with the SG  potential (circles) as a function of the cluster size at $T$ = 1 K.  is presented in Fig \ref{rhosph2_1} \cite{noterhos}. Clusters  of size $N <$ 22  are entirely superfluid (or nearly so, at least within the precision of our calculations), and display a temperature dependence of the superfluid fraction similar to that shown in Fig. \ref{18_T} for $N$=18. 

An interesting, non-monotonic trend is observed (Fig. \ref{rhosph2_1}) for the superfluid fraction at low $T$ of clusters of 22 $\le$ $N$ $\le$ 30 molecules.  The value of  $\rho_S(T$=1 K), is close to unity for $N$=22, and drops to  a local minimum if a single molecule is added (i.e., for $N$=23).
It then rises again to approximately 85\%, if another molecule is added ($N$=24), and remains relatively large  for $N$=25. The addition of another molecule, from $N$=25 to $N$=26, again causes an abrupt drop of  $\rho_S$, to less than 0.1. On adding one more molecule, $\rho_S$ becomes again significant (approximately 25\%), but drops sharply once more at $N$=29; it remains small, while still featuring noticeable oscillations, for greater  $N$ values.

Generally speaking, as the number  $N$ of particles increases, the physics of a cluster ought to approach that of the bulk; in its bulk phase at low $T$, \ph2 is an insulating (non-superfluid) crystal, i.e.,  its character is completely different from that of small clusters (i.e., $N <$  22), which remain liquid-like and are entirely superfluid. Data in Fig.  \ref{rhosph2_1} show how the evolution from liquid- to solid-like does not occur continuously, i.e.,  bulk properties do not gradually emerge when $N$ increases.

In particular, we interpret the peculiar behavior of  $\rho_S$  observed for 22 $\le$ $N$ $\le$ 30,  as due to alternating superfluid (liquid-like) or insulating (solid-like) character of the clusters. It seems plausible that drastic changes of the superfluid fraction, occurring upon adding just one molecule, ought to be directly connected with structural changes.
\begin{figure}
\begin{tabular}{cc}
\centerline{\includegraphics[scale=0.5]{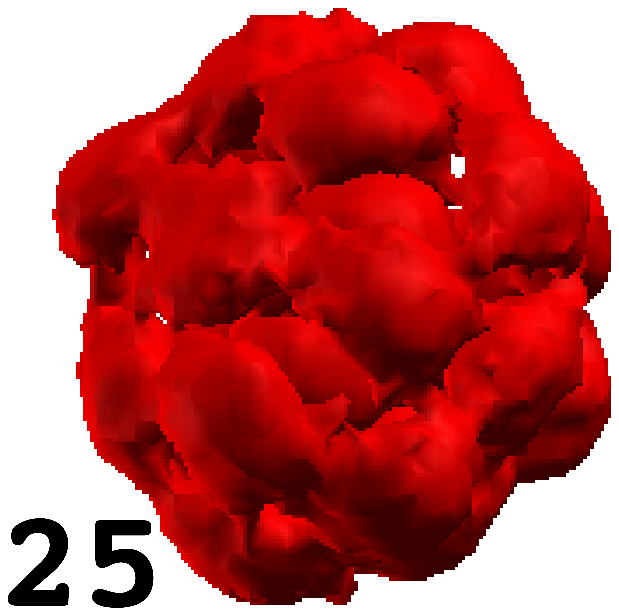}}  \\
\centerline{\includegraphics[scale=0.5]{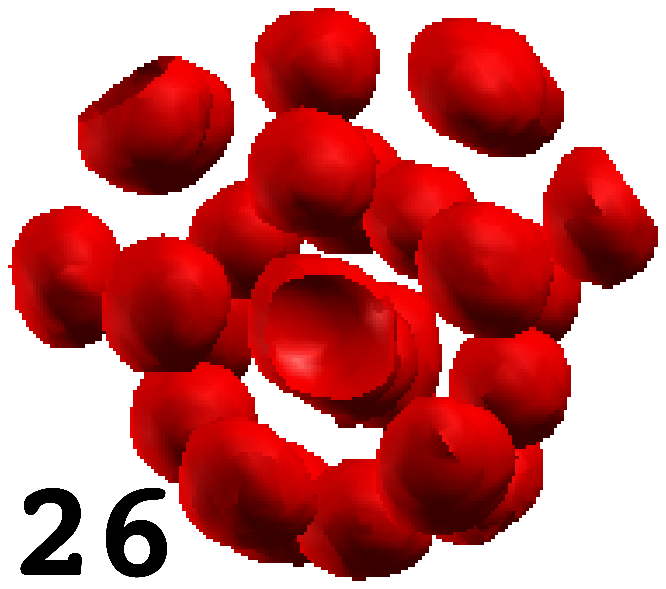}} \\
 \end{tabular}
 \caption{(color online). Three dimensional representation of a cluster comprising 25 and 26 \ph2 molecules at a temperature of 1 K. Even though the information provided by this kind of figure is purely qualitative, the structural difference is evident: (\ph2)$_{25}$ is essentially liquid-like (molecules are higly delocalized), while (\ph2)$_{26}$ is ``solid-like" (molecules are more localized and clearly distinguishable).}
 \label{25_a_26_a}
  \end{figure}  

\begin{figure}
\includegraphics[scale=0.32, angle=-90]{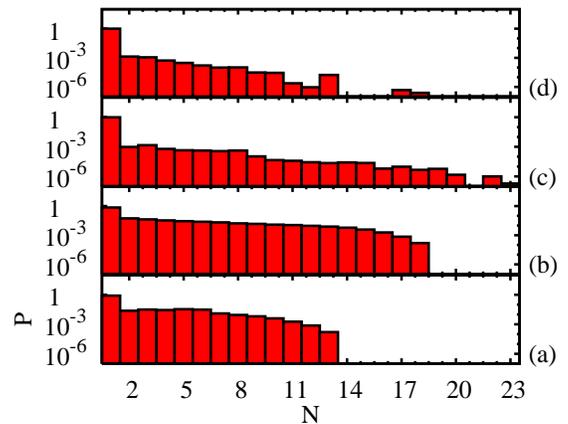}
\caption{(color online). Statistics of permutation cycles involving $N$ molecules for the clusters 
(\ph2)$_{13}$ (a), (\ph2)$_{18}$ (b), (\ph2)$_{26}$ (c) (\ph2)$_{33}$ (d) at $T$=1 K.}
\label{ph2pcycle}
\end{figure}

In order to illustrate   this point, we show  in Fig. \ref{25_a_26_a} graphical representations of the three-dimensional structure of  a  \ph2 cluster consisting of 25 and 26 molecules at $T$=1 K; these figures are produced as explained in Ref. \onlinecite{saverio05}.
The cluster (\ph2)$_{26}$ displays remarkably solid-like properties. Its structure consists of three rings of five molecules, with  four other molecules  linearly arranged along the axes of the rings, while   the remaining seven molecules form an outer shell.  Although their position is smeared by zero-point fluctuations, the various molecules in the cluster can be clearly identified, indicating that they enjoy a fairly high degree of spatial localization; consequently, exchanges among different molecules are highly suppressed (though not completely absent, as shown in Fig. \ref{ph2pcycle}), and the superfluid response is weak.

Conversely, the molecules in the cluster (\ph2)$_{25}$ cannot be clearly identified, and the entire system appears amorphous. Because of their pronounced delocalization, molecules have a strong propensity to be involved in quantum exchanges, hence the large superfluid response observed. 
The physics is reminiscent of that of the insulating-superfluid quantum phase transition observed in lattice models of hard core bosons, where a superfluid, non-crystalline phase appears as the system is driven away from commensuration (where it is an insulator) \cite{batrouni}.

Figure \ref{ph2pcycle} shows  the statistics of exchange cycles involving $N$ molecules for the 
clusters (from bottom to top) (\ph2)$_{13}$, (\ph2)$_{18}$, (\ph2)$_{26}$ and (\ph2)$_{33}$ at a temperature $T$=1 K. Long exchange cycles are known to underlie superfluidity. The value of the superfluid fraction of the smaller clusters (i.e., $N$=13 and 18) is close to 1, and exchanges occur involving up to $N$ molecules, with significant frequency. On the other hand, in the case of  (\ph2)$_{26}$ and (\ph2)$_{33}$ $\rho_S$ is less then 0.1; remarkably, however, exchange cycles involving as many as 23 and 18 molecules respectively have finite, albeit small, statistical weight. This observation is consistent with that made in Ref. \onlinecite{scharf92} (based on a PIMC simulation which did not explicitly include exchanges).  

The superfluid fraction computed with the Buck potential (boxes in Fig. \ref{rhosph2_1}) is in excellent agreement with that obtained with the SG, for those clusters that are liquid-like (i.e., with $N\le$ 22). Quantitative differences (i.e., lower values of $\rho_S$ obtained with the Buck potential, as shown in Fig. \ref{rhosph2_1}) occur for bigger clusters (which generally display more solid-like behavior), due to the more attractive character of the Buck potential, which results in greater molecular localization.  However, the observed trend is qualitatively identical.

\subsubsection{Quantum Melting}
\begin{figure}[]
\centerline{\includegraphics[scale=0.32,angle=-90 ]{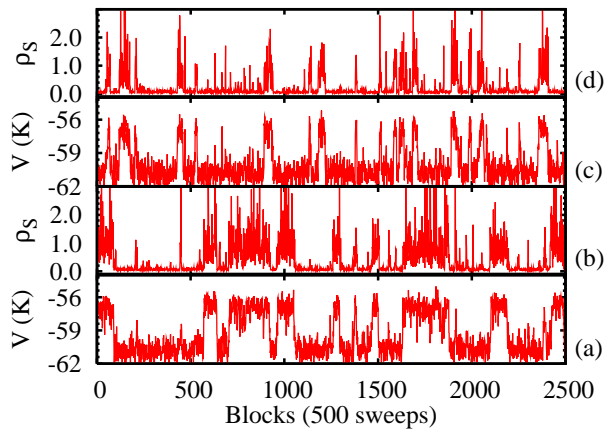}}
\caption{(color online).  Potential energy per molecule and superfluid fraction  observed during a typical Monte Carlo run (see text) for a cluster of $N$=23 \ph2 molecules at $T$=1 K (panels a and b) and $T$=1.4 K (panels c and d). The coexistence of two phases can be easily recognized since the averages of $\rho_S$ and $V$ simultaneously switch between high (liquid-like  superfluid phase) and low (solid-like insulating phase) values.  The liquid-like superfluid phase becomes dominant  as $T$ is lowered.}
 \label{qm23}
\end{figure}

Some clusters in the size range 22 $\le N \le$ 30, feature a fascinating behavior in the temperature range explored in this work; we discuss in detail the case of (\ph2)$_{23}$, for which the superfluid fraction takes a local minimum.
Fig. \ref{qm23} shows the values of the superfluid fraction and of the potential energy per molecule, recorded during a typical Monte Carlo run for a cluster of 23 \ph2 molecules at $T$=1.4 K and $T$=1 K. In particular, we show consecutive block averages of $\rho_S$ and $V$,  each block consisting of 500 sweeps \cite{notesweep}. 

Despite the large fluctuations affecting the values of $\rho_S$,  two different regimes can be easily identified: one in which the superfluid fraction is high, with an average value close to 1, and the other characterized by low values of  $\rho_S$, with an average value close to zero. The potential energy, correspondingly, takes on high (low) values when   $\rho_S$ is large (small). While the behavior of the superfluid fraction is consistent with coexistence of two phases, characterized by large and small superfluid response, the information given by the potential energy suggests that these two phases have also liquid-like and solid-like properties.  For, the switching of the average value of the potential energy between two different regimes, separated by some $\sim$ 6 K, can be interpreted as due to the system visiting relatively ordered, solid-like, insulating configurations (characterized by low potential energy), and disordered, liquid-like, superfluid ones.  The  coexistence between two distinct (disordered and ordered) phases is consistent with the observation made in Ref. \cite{(kunz93} for classical clusters. 

On decreasing the temperature from $T$=1.4 K (panels {\it c} and {\it d} of Fig. \ref{qm23})  to $T$=1 K (panels {\it a} and {\it b} of the same figure), the superfluid (liquid-like) phase (i.e., large values of $\rho_S$ and $V$ ) is observed during a greater fraction of the simulation time, and   becomes dominant as $T$  approaches 0 K; the cluster therefore ``melts" at low $T$. This process is ostensibly induced by zero point motion, and the ensuing exchanges of molecules, whose importance increases as $T$ is lowered. In this sense, one could state that melting is driven by Bose statistics, i.e., it is associated to the energy contribution due to quantum exchanges.

\begin{figure}[]
\centerline{\includegraphics[scale=0.32,angle=-90 ]{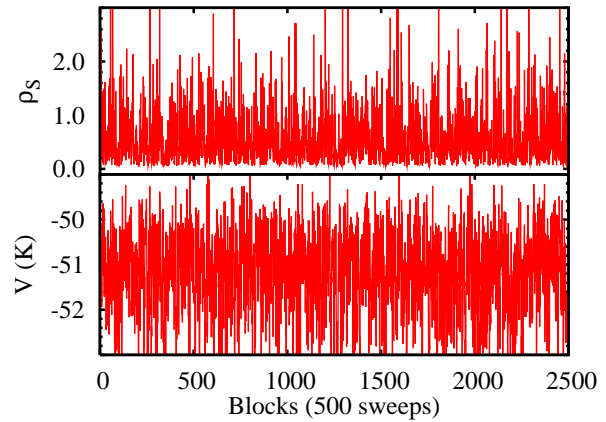}}
\caption{(color online).  Potential energy per molecule and superfluid fraction  recorded during a typical Monte Carlo run  for a cluster of $N$=18 \ph2 molecules at  $T$=2 K. In this case, $\rho_S$ and $V$ simply oscillate around their average values (approximately 0.6 for $\rho_S$) without featuring the clear, simultaneous ``jumps" observed in Fig \ref{qm23}. This system is found to be liquid-like in the range of temperature considered in this work.}
 \label{18}
\end{figure}

On the other hand, when $T$ is sufficiently high, quantum exchanges are suppressed and the system ``freezes" in a solid-like structure. Evidence of such {\it quantum melting} has also been found  for (\ph2)$_{27}$.  Other clusters in the size range 22 $\le$ $N$ $\le$ 30 presumably display the same physics, albeit at lower $T$. It is important to stress that the behavior illustrated in Fig. \ref{qm23} is different than that of a cluster that is liquid-like, and simply not 100\% superfluid; indeed, in such a case superfluid fraction and potential energy merely fluctuate around their average values, with no evidence of the switching displayed in Fig. \ref{qm23}. This is precisely what we find  for clusters of size $N <$  22 , which are liquid-like at all temperatures, with superfluid (normal) component growing at low (high) $T$ (see Fig. \ref{18}). 

\begin{figure}
\centerline{\includegraphics[scale=0.33,angle=-90]{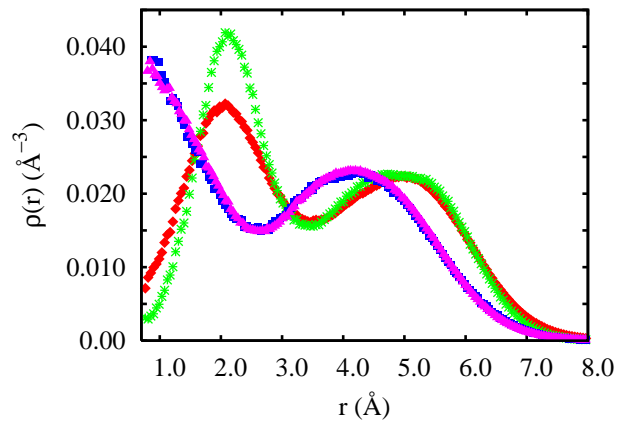}}
\caption{(color online).  Radial density profiles for a \ph2 cluster with $N$=18  [ $T$=0.75 K (boxes) and $T$=2 K (triangles)] and $N$=23 [$T$ = 0.75 K (diamonds) and $T$=2 K (stars)] \ph2  molecules. Statistical errors, not shown for clarity, are of the order of 5$\times10^{-4}$ \Am3 or less.}
 \label{23_18_T}
\end{figure}

Our interpretation  is supported by the evolution with $T$  of the radial density profiles $\rho(r)$ of the cluster, defined with respect to its center of mass. In Fig. \ref{23_18_T}, radial density profiles for (\ph2)$_{23}$ and (\ph2)$_{18}$ at $T$=0.75 K (diamonds and boxes) as well as at $T$=2 K (stars and triangles) are shown. A cluster comprising 23 \ph2 molecules at a temperature $T$=2 K, displays a two-shell structure with one sharp peak at $r \approx$ 2\   \AA, and a second broader one at about 5.5    \AA. As $T$ is lowered to 0.75 K,   the first peak  becomes significantly broader,  and its height  decreases. Therefore, molecules in the inner shell are less localized, and the cluster is more liquid-like, with greater propensity for quantum exchanges both in the first shell, as well as between the first and the second shells. For a cluster of 18 molecules, on the other hand, density profiles stay the same, as $T$ is lowered from 2 K to 0.75 K,  featuring one particle in the center of mass [signaled by the large value of $\rho(r\to 0)$] and an outer shell, separated by a shallow minimum (indicative of liquid-like structure). The value of the superfluid fraction of (\ph2)$_{18}$ increases from about 0.57 to roughly 1 when $T$ goes from 2 K to 0.75 K; however, this change is hardly reflected in the cluster structure. Quite differently, the increase of $\rho_S$ in a cluster of 23 molecule, observed in the same temperature range, reflects a substantial  change in the structure, due to quantum melting (as shown in Fig. \ref{23_18_T}). This phenomenon clearly exemplifies  the importance of quantum effects on the structural properties of small systems,  which has been also investigated in clusters of neon, which display moderate quantum character  \cite{predescu05, calvo01, predescu01, frantsuzov06}.

We conclude by stressing once again that, even though the results presented in this section are obtained using the SG potential, similar physics is observed if the more attractive Buck potential is adopted. The only quantitative changes are the values of the superfluid fraction, which are generally lower than those computed with the SG interaction for a given $T$. However, quantum melting is observed in clusters of the same size, even if the potential due to Buck is used, i.e., it is broadly independent of the details of the interaction.

\subsection{\od2 clusters}
\label{od2results}
Next, we discuss the properties of  \od2 clusters; here too, we regard \od2 molecules as bosons of spin zero \cite{notespin}, with a mass twice that of \ph2 molecules. We computed energy, superfluid fraction and radial density profiles for clusters of size $N$ ranging from 3 to 20 molecules at 0.5 K $\le$ $T$ $\le$ 2.0 K.

\subsubsection{Energetics}
\begin{figure}
\centerline{\includegraphics[scale=0.32,angle=-90 ]{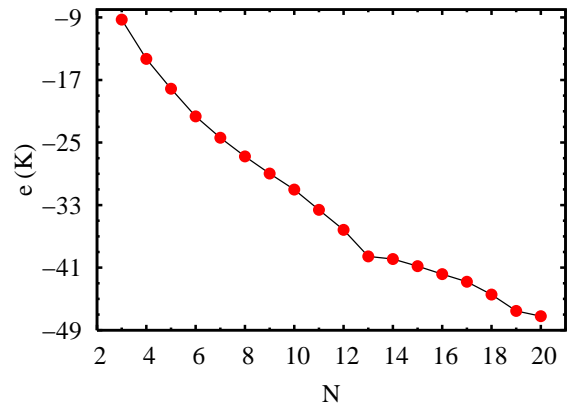}}
\caption{(color online). Energy per \od2 molecules $e$ versus cluster size $N$ at a temperature of 0.5 K.   Statistical errors are smaller  than the symbol size. Solid line is only a guide to the eye.}
 \label{energy_05_od2}
\end{figure}

\begin{figure}[t]
\includegraphics[scale=0.32, angle=-90]{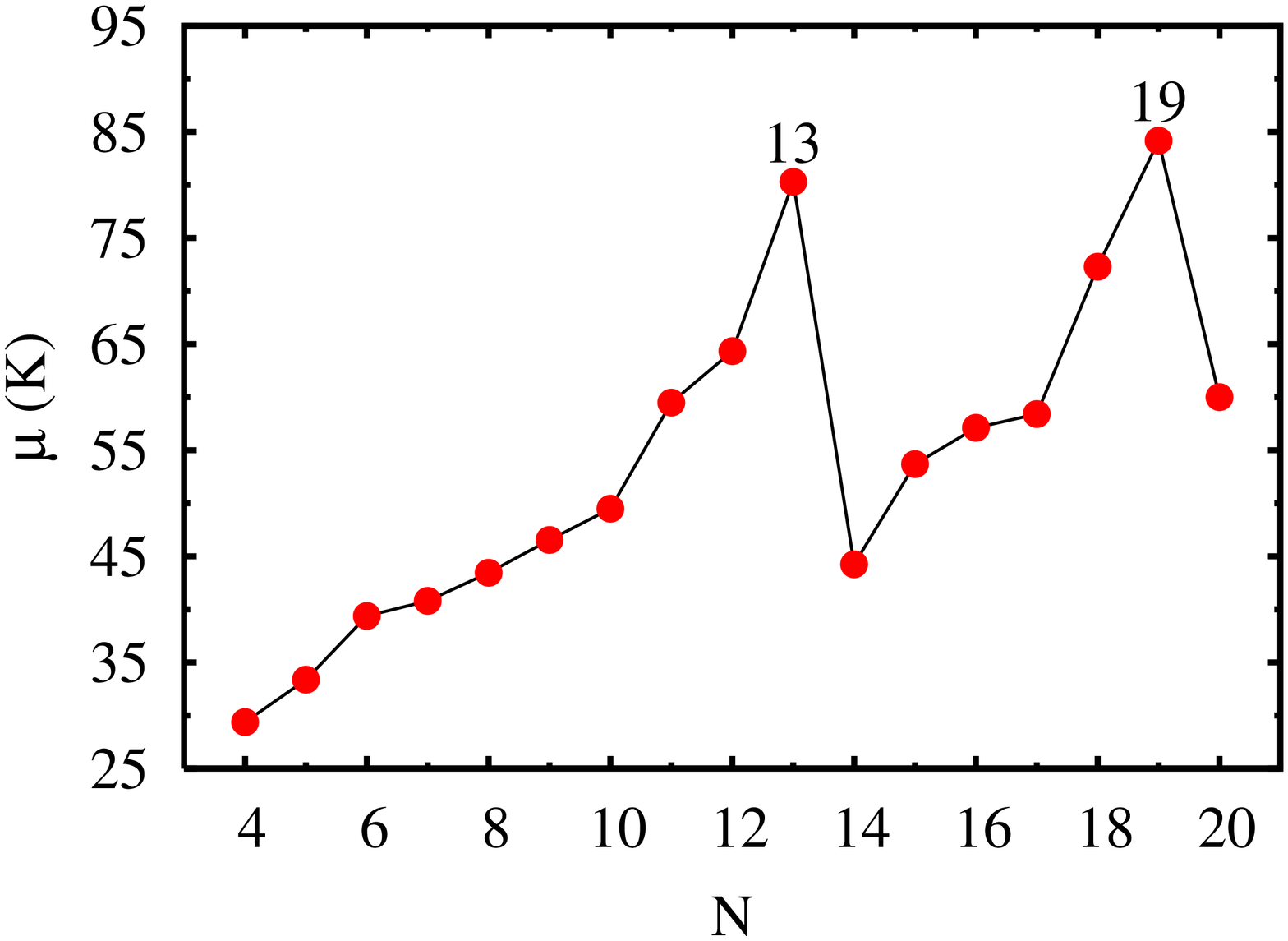}
\caption{(color online). Chemical potential of \od2 clusters versus cluster size $N$. When not shown, statistical errors are of the order of, or smaller than the symbol size. Solid line is only a guide to the eye. Numbers refer to stable (``magic") clusters.}
\label{cmptlod2}
\end{figure}

The total energy per \od2 molecule as a function of the cluster size, at $T$=0.5 K is shown in Fig. \ref{energy_05_od2}.  For $N$=3, the energy has a value close to $-$9 K, which smoothly decreases with $N$,  reaching a value of roughly $-$36 K for $N$=12. On adding one more molecule, the energy drops to about $-$39.5 K ($N$=13), and decreases more slowly for  $N$ up to 17. For greater values of $N$ an important drop of the energy is still observed at $N$=19.  As in the previously discussed case of \ph2 clusters, a fit of our data using Eq. \ref{fit} is difficult, and does not allow an accurate estimate of the bulk energy; moreover, the  chemical potential, (see Fig. \ref{cmptlod2}) displaying  sharp local maxima for $N$=13 and 19  confirms the existence of  ``magic'' sizes, corresponding to  highly stable clusters  [i.e., (\od2)$_{13}$ and (\od2)$_{19}$] 

\subsubsection{Superfluidity}

 \begin{figure}[]
\includegraphics[scale=0.32, angle=-90]{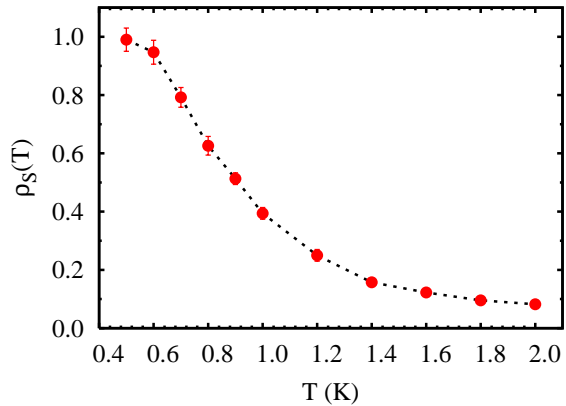}
\caption{(color online). Temperature dependence of the superfluid fraction $\rho_S(T)$ for a clusters of 11 \od2 molecules  Dotted line is only a guide  to the eye. When not shown, statistical errors are smaller than the symbol size.}
\label{11_T}
\end{figure}

\begin{figure}
\includegraphics[scale=0.32, angle=-90]{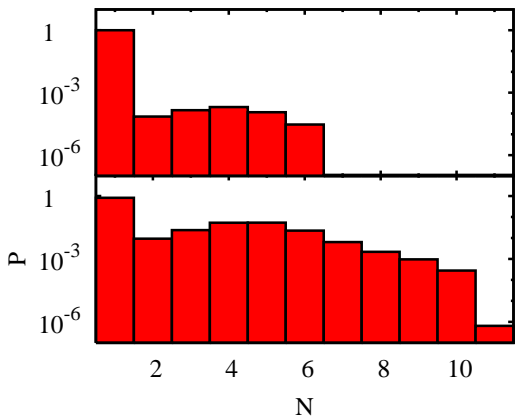}
\caption{(color online). Statistics of permutation cycles involving $N$ molecules for the cluster
(\od2)$_{11}$  at $T$=0.5 K (lower panel) and $T$=2 K (upper panel).}
\label{od2pcycle}
\end{figure}

Fig. \ref{11_T} shows $\rho_S(T)$ for a cluster of 11 \od2 molecules. The monotonically decreasing trend is qualitatively similar to that in Fig. \ref{18_T} for (\ph2)$_{18}$.  However, the superfluid fraction of the cluster (\od2)$_{11}$ decreases more rapidly with $T$. In particular, it is  approximately 1 at $T \le$ 0.5 K, and suddenly drops to a value $\simeq$ 0.39 (corresponding to roughly 4 molecules in the superfluid phase) when $T$ increases of 0.5 K. At higher temperature, the superfluid fraction becomes essentially zero (in the sense explained in the previous section) for  $T \simeq$ 1.4 K, at which the average number of molecules in the superfluid phase is less than two. It is important to note that, while the cluster (\ph2)$_{18}$ is almost  60\% superfluid at $T$=2 K (Fig. \ref{18_T}),  (\od2)$_{11}$ at the same temperature is non-superfluid, and the probability of observing permutation cycles comprising more than 3 \od2 molecules is less than 5 $\times$ 10$^{-4}$. The statistics of many-particle permutations in the system (\od2)$_{11}$ at $T$=0.5 K (lower panel) and 2 K (upper panel)  shown in Fig. \ref{od2pcycle} points out how, at high $T$, long exchange cycles disappear, as clusters behave more classically. 

\begin{figure}
\centerline{\includegraphics[scale=0.32, angle=-90]{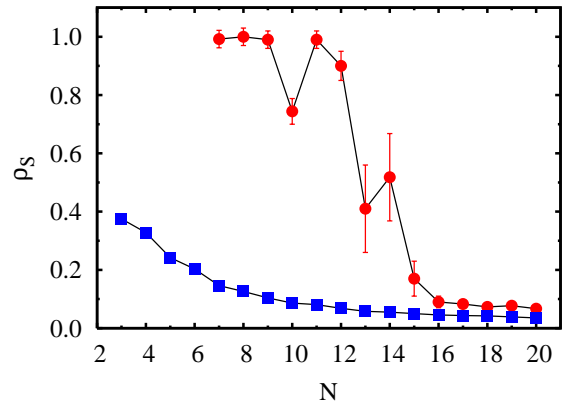}}
\caption{(color online). Superfluid fraction of \od2 clusters versus cluster size $N$, at $T$=0.5 K (circles)  and $T$ = 2 K (boxes).  When not shown, statistical errors are of the order of, or smaller than the symbol size. Solid lines are only  guides to the eye.}
\label{rhosod2}
\end{figure}

The size dependence of the superfluid fraction at two different temperatures  ($T$=0.5  and 2 K) is shown in Fig. \ref{rhosod2}. At $T$=0.5 K,  small \od2 clusters  ($N \le$ 9 ) are entirely superfluid, while for larger values of $N$  the trend of $\rho_S$ is consistent with a noticeable dependence on the cluster size, similarly to what observed for \ph2 clusters with 22$\le N\le$ 30. A local minimum is observed for $N$=10; then the superfliuid fraction grows to a value close to 1 for $N$=11, and remains relatively large at $N$=12.

Large fluctuations observed during lengthy simulations, render the determination of the value of $\rho_S$ quite difficult for clusters greater than 11 \od2 molecules. Nevertheless, we can conclude from our data that clusters of up to 14 molecules still feature a significant superluid response, which instead is close to zero for larger clusters. As expected, at higher temperatures $\rho_S$ becomes progressively smaller and many-particle permutation cycles are almost absent. 

The behavior shown in Fig. \ref{qm23}, indicative of quantum melting, has not been observed for \od2 clusters, in the temperature range explored in this work.  We expect it to be observable at lower temperatures. 

\subsubsection{Structure}
\begin{figure}
\centerline{\includegraphics[scale=0.33,angle=-90 ]{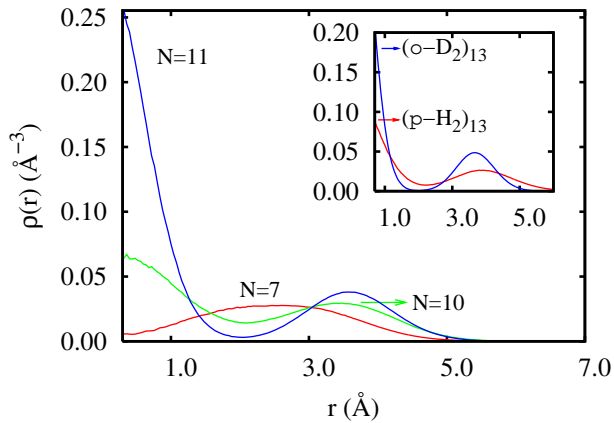}}
\caption{(color online).  Radial density profiles for  \od2 clusters with $N$=7, 10 and 11, at $T$ = 0.5 K. The inset shows the radial density profiles of (\od2)$_{13}$ and  (\ph2)$_{13}$ at the same temperature. Statistical errors, omitted  for clarity, are of the order of 5$\times10^{-4}$ \Am3 or less.}
 \label{d2rad}
\end{figure}

Radial density profiles of the clusters (\od2)$_7$, (\od2)$_{10}$ and (\od2)$_{11}$  at $T$=0.5 K are shown in Fig. \ref{d2rad}. The cluster (\od2)$_7$ features a single-shell structure, with a broad peak at about 2.5 \AA, consistent with a high degree of delocalization of the molecules.  For $N$=10, the probability of having the center of the system occupied by a molecule becomes significant, and the peak is shifted to higher distances by roughly 1 \AA.
The shallow minimum which appears at $r \simeq$ 2 \AA, suggests that exchanges are frequent, and not restricted to molecules in the first shell, i.e., they  involve also the molecule in the center of the cluster.

Important structural differences characterize (\od2)$_{11}$; as shown by the small value of the radial
density at the minimum,
and by the increased height of the first peak in Fig. \ref{d2rad}, the molecule in the center of the cluster  is strongly localized [see also the high value of $\rho(r\rightarrow 0)$], and scarcely participates  to multi-particle exchanges. Remarkably, the   superfluid fraction of all of these systems is large, close to 1 for $N$=7 and 11. The inset of Fig. \ref{d2rad} compares radial density profiles of  (\od2)$_{13}$ and (\ph2)$_{13}$ at $T$=0.5 K; the effect of mass difference on the structural properties is evident. Indeed, the density profile of (\ph2)$_{13}$ reflects the liquid-like  nature of such a system, whereas in the case of (\od2)$_{13}$, structural details (i.e., pronounced first shell peak and the small value of the radial density at the minimum) are consistent with a more orderly (i.e., solid-like) structure. 
\begin{figure}
\begin{tabular}{cc}
\centerline{\includegraphics[scale=0.5]{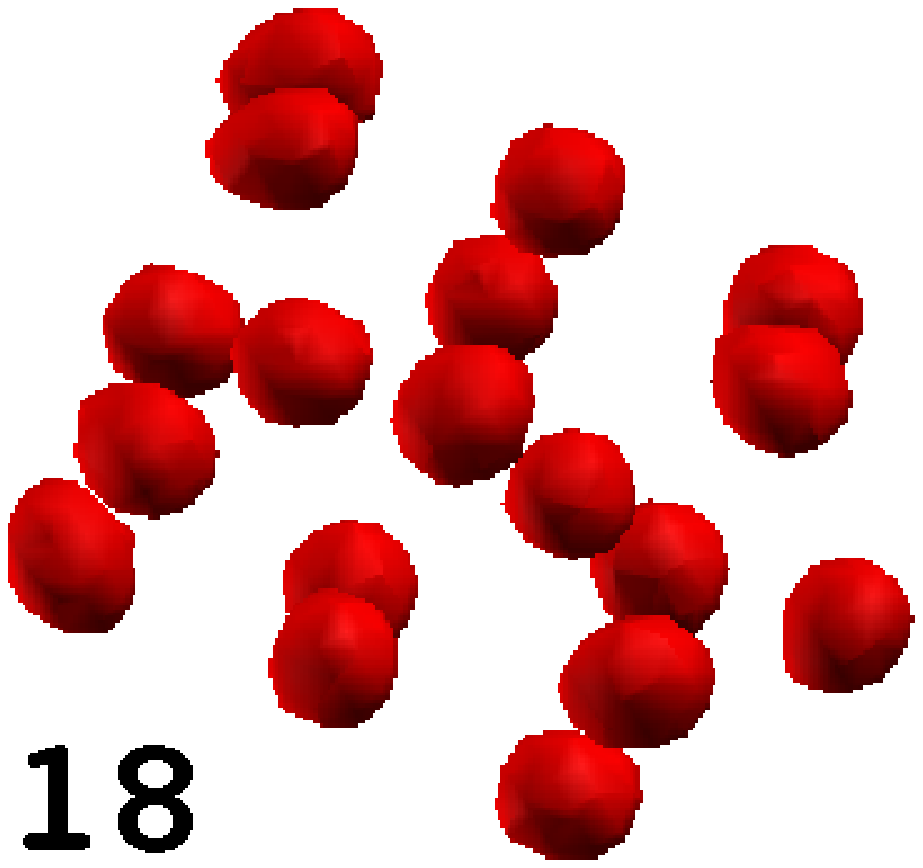}}  \\
\centerline{\includegraphics[scale=0.5]{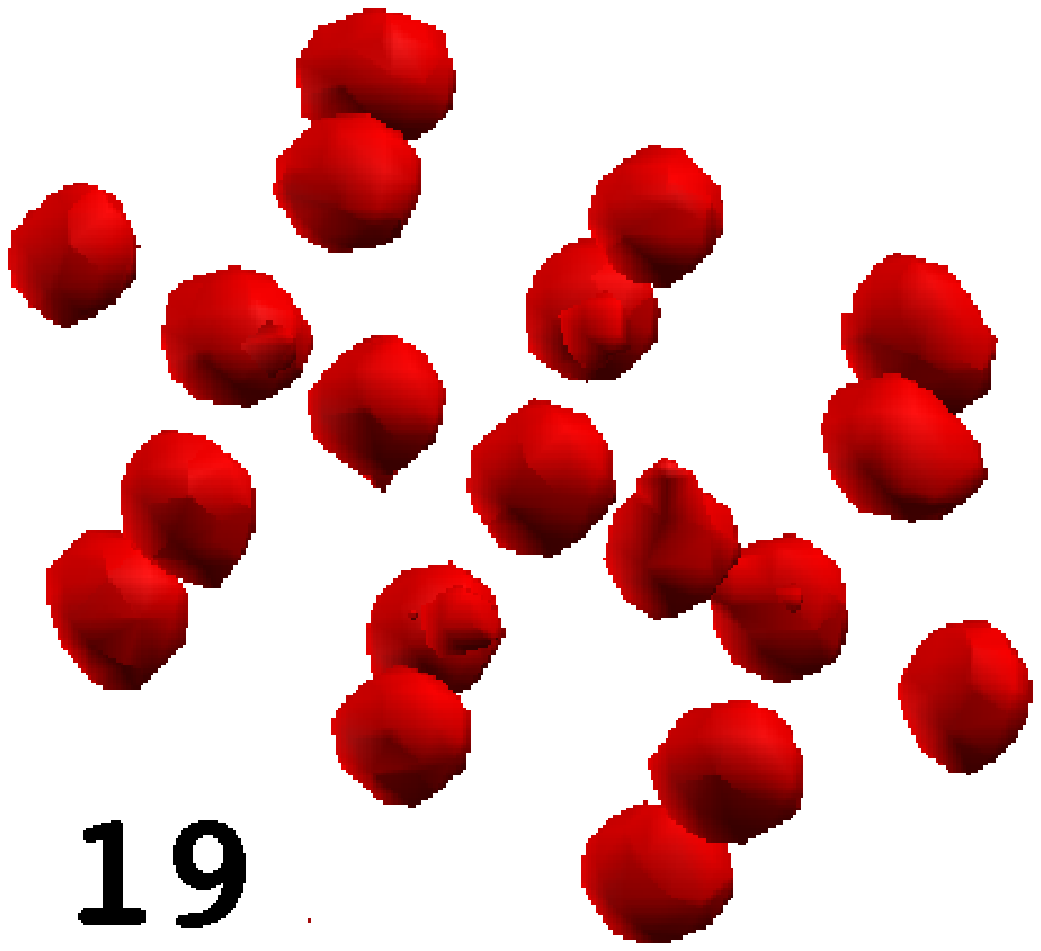}} \\
 \end{tabular}
 \caption{(color online). Three-dimensional representation of  clusters comprising 18 and 19 \od2 molecules, at a temperature of 0.5 K. Both systems have solid-like properties, as molecules enjoy a high degree of localization. However, (\od2)$_{19}$ features greater stability, due to its more symmetric structure (see text).}
 \label{18_19}
  \end{figure}  

It has been already pointed out how local maxima of the chemical potential correspond to magic clusters, featuring greater stability than others (see Fig. \ref{cmptlod2}). Here, we discuss in detail the case of (\od2)$_{19}$. In Fig, \ref{18_19} we show the three-dimensional structure of clusters comprising $N$=18 and 19 \od2 molecules, at $T$=0.5 K. The cluster (\od2)$_{18}$ is made of three rings of five molecules, with the remaining three molecules linearly arranged along the axes of the rings. On adding a single molecule ($N$=19), which also positions itself along the axes of the rings, the structure becomes symmetric with respect to the plane of the central ring. Since both  (\od2)$_{18}$ and (\od2)$_{19}$ are solid-like and insulating (i.e., molecules are clearly distinguishable and the value of $\rho_S$ is less than 0.1), the sharp maximum of the chemical potential at $N$=19  appears not to be related to noticeable structural changes (i.e., from more liquid- to solid-like, as observed  for \ph2 when $N$ increases from 25 to 26), but  rather to the pronounced symmetry of this cluster.
 
\section{CONCLUSIONS}
We have studied the low temperature physical properties of \ph2 and \od2 clusters, of size up to 40 and 20 molecules respectively, using an accurate numerical method. The superfluid fraction of (\ph2)$_N$ clusters displays a remarkable dependence on the size $N$  for 22  $\le$ $N$ $\le$ 30 pointing out how the bulk phase emerges non-monotonically. Coexistence of liquid-like (superfluid)  and solid-like (insulating) phases has been shown in some clusters in this size range; the dominance of the first phase at low $T$ indicates that these systems melt as a result of the zero point motion and quantum exchanges, freezing instead at high $T$. 
Superfluid behavior has been also found in \od2 clusters of size up to $N$=14. The heavier mass of the \od2 molecules is responsible for the suppression of $\rho_S$ in larger clusters.
Some clusters, featuring specific  numbers of particles, are characterized by greater stability; the fact that the ``magic number"  $N$=13 is present for both \ph2 and \od2 clusters, suggests that its stability is associated with high symmetry, i.e., it is related to the potential energy. These structures, however, need not necessarily be solid-like, but rather simply have filled coordination shells.
Given the intense level of experimental activity in this field, it is to be hoped that some of the predictions made in this work will be soon tested experimentally.

\section*{ACKNOWLEDGMENTS}
This work was supported  by the Natural Science and Engineering Research Council of Canada under research grant 121210893. Simulations were performed on the Mammouth cluster at University of Sherbrooke (Qu\`ebec, Canada). 

{\it Note added.} Recently we became aware of a manuscript by Khairallah {\it et al.} \cite{khairallah}, also investigating superfluid and structural properties of para-hydrogen clusters by path integral Monte Carlo simulations. Although some of the model interactions utilized therein are not the same as in this work, and therefore a direct comparison is problematic, their results appear in qualitative agreement with ours.


\begin{thebibliography}{wide-label}
\bibitem{noteliquid} Strictly speaking, of course, ``liquid", ``solid" etc. are only meaningful definitions for bulk phases. In this manuscript, as is customary in the literature when speaking of clusters, they are used somewhat loosely, with reference to the presence of order, reminiscent of that of a crystal, on a scale comparable to the size of the cluster.
\bibitem{notepbc}
The size of the simulation cell turns out to be at least twice the size of the biggest cluster considered in this work, which renders PBC (used for convenience only)  immaterial.
\bibitem{notextra}  We have computed energetics of several clusters at $T$ lower than 1 K, and estimate the maximum difference between our results at $T$=1 K and those extrapolated to $T$=0 to be less than 0.3 K.
\bibitem{noterhos} The results presented here are consistent with those reported in Ref. \cite{noi06}, once statistical uncertainties are properly taken into account. The estimates of $\rho_S$ presented in this work benefit from much longer computer runs than those of Ref. \cite{noi06}, hence their smaller statistical uncertainties.
\bibitem{notesweep}We define a {\it sweep} as a sampling cycle which starts with the opening and ends with the closing of a worm (for details, see Refs. \onlinecite {MBworm}).
\end{thebibliography}
\end{document}